\documentclass[fleqn,usenatbib, useAMS]{mnras}

\usepackage{graphicx}	
\usepackage{amsmath}	
\usepackage{amssymb}	
\usepackage{multicol}        
\usepackage{bm}		
\usepackage{pdflscape}	
\usepackage{flafter}
\usepackage{lscape}
\usepackage{times}
\usepackage{lmodern} 
\usepackage[T1]{fontenc}
\usepackage[utf8]{inputenc}
\usepackage{hyperref} 
\def\be{\begin{equation}}
\def\ee{\end{equation}}
\def\bea{\begin{eqnarray*}}
\def\eea{\end{eqnarray*}}
\newcommand{\degree}{\ensuremath{^\circ}}

\def\url#1{\expandafter\string\csname #1\endcasname}
\def\H2{$H_2$}

\def\kms{km s$^{-1}$}

\def\Herschel{\textit{Herschel}}
\def\Planck{\textit{Planck}}

\def\LMT{\textit{LMT}}
\def\GBT{\textit{GBT}}

\def\mnras{MNRAS}

\def\aap{A\&A}
\def\araa{ARA\&A}
\def\aj{AJ}
\def\apjl{ApJ}
\def\apj{ApJ}
\def\apjs{ApJS}

\def\pasp{PASP}
\def\nat{Nature}
\def\physrep{Physics Reports}

\begin{document}
\title[Total Molecular Gas of \Planck\ -- \Herschel\ HyLIRGs
]{Total Molecular Gas Masses of \Planck\ -- \Herschel\ Selected Strongly Lensed Hyper Luminous Infrared Galaxies}

\author[K. C.~Harrington et al.]{K.~C. ~Harrington$^{1,2}$\thanks{E-mail: kharring@astro.uni-bonn.de}
M. S.~Yun$^3$,
B.~Magnelli$^2$,
D.T. ~Frayer$^4$,
\newauthor
A.~Karim$^2$,
A. ~Wei{\ss}$^1$,
D. ~Riechers$^5$,
E. F. ~Jim{\'e}nez-Andrade$^{1,2}$,
D. ~Berman$^3$,
\newauthor
J. ~Lowenthal$^6$, and
F.~Bertoldi$^2$
\\
$^1$Max-Planck-Institut f\"{u}r Radioastronomie, Auf dem H\"{u}gel 69, 53121 
Bonn, Germany\\
$^2$Argelander Institut f\"{u}r Astronomie, Auf dem H\"{u}gel 71, 53121 Bonn, 
Germany\\
$^3$Department of Astronomy, University of Massachusetts, 619E Lederle Grad 
Research Tower, 710 N. Pleasant Street, Amherst, MA 01003, USA\\
$^4$Green Bank Observatory, 155 Observatory Rd., Green Bank, West Virginia 24944, USA\\
$^5$Department of Astronomy, Cornell University, Space Sciences Building, Ithaca, NY 14853, USA\\
$^6$Department of Astronomy, Smith College,  Northampton, MA 01063, USA\\
}

\date{\today}
\pagerange{\pageref{firstpage}--\pageref{lastpage}} \pubyear{2017}
\maketitle
\label{firstpage}

\begin{abstract}
We report the detection of CO($1-0$) line emission from seven  \Planck\ and \Herschel\
selected hyper luminous (${\rm L_{\rm IR (8-1000\micron)} > 10^{13} L_{\odot} }$) infrared galaxies
with the Green Bank Telescope (\GBT). 
CO($1-0$) measurements are a vital tool to trace the bulk molecular gas mass across all redshifts.
Our results place tight constraints on the total gas content of these most apparently luminous high-$z$ star-forming galaxies
(apparent IR luminosities of $L_{\rm IR} > 10^{13-14} {\rm L_{\odot}} $), while we
confirm their predetermined redshifts measured using the Large Millimeter Telescope, \LMT\ 
($ z_{\rm CO} = 1.33 - 3.26$).
The CO($1-0$) lines show similar profiles as compared to J$_{\rm up} = 2-4$ transitions previously observed with the 
\LMT. We report enhanced 
infrared to CO line luminosity ratios of $ < L_{\rm IR} /  L'_{\rm CO(1-0)} > = 110 \pm 22 
\, 
{\rm L_{\odot} (K \, km \, s^{-1} \, pc^{-2})^{-1} } $
compared 
to normal star-forming galaxies, yet similar to those of well-studied 
IR-luminous galaxies at high-$z$. We find average brightness temperature ratios of ${\rm < r_{21} > = 0.93}$ (2 sources), ${\rm < r_{31} > = 
0.34}$ (5 sources), and ${\rm < r_{41} > =  0.18}$ (1 source). The ${\rm r_{31}}$ and ${\rm r_{41}}$ values are roughly half the 
average values for SMGs. We estimate the total gas mass content as ${\rm \mu 
M_{H2} = (0.9 - 27.2 ) \times 10^{11} (\alpha_{\rm CO}/0.8) M_{\odot}}$, 
where ${\rm \mu}$ is the magnification factor and $\alpha_{\rm CO}$ is the CO line luminosity to molecular hydrogen gas mass conversion factor. The rapid 
gas depletion times, ${\rm < \tau_{\rm depl} > = 80}$ Myr, reveal vigorous
starburst activity, and contrast the 
Gyr depletion timescales observed in local, normal star-forming galaxies. 

\end{abstract}

\section{Introduction}
Most of the stars in the local universe formed out of 
tremendous cool gas reservoirs ($M_{\rm gas} \sim 10^{10-11} {\rm M {\odot} }$, $T \sim10-100$ K) 
in the interstellar medium (ISM) of high redshift (${\rm 1 < z < 
3.5}$) galaxies with intense star-formation (SF) \citep{carilli13,madau14}. 


Massive, dusty star forming galaxies at high-$z$ (DSFGs; $M_{\rm dust} \sim 10^{8-9} {\rm M_{ \odot} }$) are typically gas-rich galaxies
selected via their bright observed (sub)-mm fluxes (also known as SMGs). 
The rest-frame far-IR (FIR)-mm luminosity associated with the thermal dust emission 
\citep{efstathiou00, johnson13} (re-radiated far-UV radiation) traces the total star-formation (SF) activity, while the extreme 
star-formation 
rates (SFRs) in these
IR luminous galaxies are likely due to a sustained supply of cool gas from the intergalactic medium (IGM). 
The dense molecular ingredients of the ISM thereby limits the timescale for extended starburst (SB) activity, with short-lived SB episodes of 10's-100's of
Myr. These are believed to often include gas-rich mergers that induce star formation via tidal torques,
which drive gas infall and subsequent collapse \citep{hern92}. The most active SB galaxies
at $z \sim 2$ contribute key insights into galaxy evolution and structure formation, as
their massive gas reservoirs play a key role in the bulk stellar mass
growth in their ISM environments, and as a result are believed to be the progenitors to massive
elliptical/spheroidal galaxies and clusters at low-$z$ \citep{casey14}. 

The
SMG population can be accounted for by major or minor-merger dominated starbursts \citep{baugh05,swinbank08} in
some semi-analytic models. Others suggest that the 
observed population is a heterogeneous mix of early and late stage major mergers and
blending of passive star-forming disc galaxies. The brightest SMGs are likely early-stage mergers,
exchanging a significant amount of molecular material for continued star formation \citep{hayward12, narayanan15}. 
SMGs typically have high gas mass fractions, ${\rm M_{gas} / M_{\star} }$, up to 
80\% \citep{carilli13}.

CO line measurements are vital for directly probing the fuel for these star-forming galaxies, i.e. the total 
molecular gas mass, at the peak of the co-moving SFR density (${\rm z \sim 2-3}$). 
The gas accretion 
history of growing dark matter (DM) halos in cosmological simulations \citep{keres05} agrees 
well with the observed evolution of the CO luminosity function, as
\citet{decarli16} find a peak redshift for CO luminous galaxies at $z \sim 2$, 
comparable to the peak of the co-moving star formation rate density. 

The CO (J$=1\rightarrow 0$) transition accounts for both the 
dense and most diffuse molecular gas, and has traditionally been calibrated to 
trace the bulk
H$_{\rm 2}$ gas mass (via collisional excitation with the H$_{\rm 2}$ gas). 
The observed CO line luminosity, L'$_{\rm CO}$, to ${\rm H_{2}}$ mass conversion 
factor, $\alpha_{\rm CO}$, 
\citep{carilli13, bolatto13}, is calibrated to this transition, making observations of CO($1-0$)
important for determining the total ${\rm H_{2}}$ mass content at high-$z$.

The number of high-$z$ 
sources with galaxy integrated CO($1-0$) detections is sparse \citep[see][]{scoville17,carilli13}, although it is accumulating 
\citep[e.g.][]{carilli02, hainline06,
riechers11a,harris12, thomson12, riechers13,fu13, aravena13, sharon16, decarli16, decarli16b, huynh17}, with approximately 
60 to date. Resolved imaging of this lowest rotational transition of 
CO \citep[e.g.][]{riechers11a,lestrade11} in high-$z$ SMGs 
indicates that the total molecular gas
can extend up to 30 kpc for merging systems.
Only the most active star forming 
sources with apparent $L_{IR} \ge10^{12-14} L_\odot$ at ${\rm z > 1}$ can be 
observed at this fundamental CO rotational transition. These apparent 
luminosities are often due to strong lensing. The strong lensing effect 
(usually with magnification factor, $\mu = 10-30X$)
\citep[e.g.][]{bussmann13, bussmann15, geach15, spilker16a}, yields shorter integration times to provide secure detections of the molecular gas 
in both strongly lensed, intrinsically bright and faint, but highly magnified, normal star-forming systems. 

The far-IR/sub-mm 
\Herschel\ Astrophysical Terahertz Large Area Survey (H-ATLAS) \citep{eales10} and the Herschel Multi-tiered Extragalactic Survey (HerMES)
 \citep{oliver12}, together covering about 650${\rm 
\deg}^{2}$, and the 2500 ${\rm \deg}^{2}$ mm South Pole 
Telescope \citep[\textit{SPT},][]{carlstrom11} have paved the way forward in 
discovering a rare population of gravitationally lensed 
DSFGs \citep[e.g.][]{negrello10, planck7, wardlow13, vieira13, weiss13, negrello17}, as well as an 
intrinsically bright, unlensed population. \citet{canameras15} and 
\citet{harrington16} have exploited the all-sky sensitivity of \Planck\ to find 
the most luminous high redshift galaxies currently known in the \Planck\ era -- all of which are gravitationally 
lensed \citep[also][]{herranz13,fu12, combes12}. 

  Here we present galaxy integrated, CO($1-0$) measurements of seven $z >$1 galaxies using the Green Bank Telescope 
(\GBT). This is a pilot study for a larger program to identify a large sample of 
extremely luminous high-$z$ SMGs identified by the all-sky \Planck\ 
survey. 
In our original pilot study leading to the sample in this work, our goal was to identify sources that have the probability 
to be gravitationally lensed given their high flux densities in the 3 SPIRE 
bands of 250, 350, 500\micron \,  \citep[e.g. the ${\rm S_{500} \, {\rm or} \, S_{350} \ge 100 \,
mJy}$][]{negrello10, ivison11}.  We have previously obtained one J$_{\rm up} = 2-4$ transition for all seven of the sources presented in this study using the 
Redshift Search Receiver (RSR) on the \LMT. The majority of these sources have apparent ${\rm \mu L_{IR} > 
10^{14.0-14.5}  L_\odot}$ making them some of the most luminous sources 
currently known \citep{harrington16, canameras15}. Our goals in this study are 
to confirm the \LMT\ CO redshift, measure the CO($1-0$) line emission to constrain our 
estimate of the ${\rm H_{2}}$ masses and begin analysing the CO spectral line energy distributions (CO SLEDs). 
In \S~2 we 
review our sample selection and previous observations described in detail in 
\citet{harrington16}, and in 
\S~3 we outline our CO($1-0$) observations using the VErsatile GBT Astronomical 
Spectrometer (VEGAS) instrument on the \GBT. Measured and derived gas 
properties 
using the CO($1-0$) line emission and supplementary \LMT\ CO data is found in 
\S~4, followed by a discussion in \S~5. Finally, we conclude our study in \S~6. 
We adopt a $\Lambda$ CDM cosmology with ${\rm H_{0} = 70 \, km s^{-1} Mpc^{-1} 
}$ with ${\rm \Omega _{m} = 0.3}$, and ${\rm \Omega _{\Lambda} = 0.7}$ 
throughout this paper.

\section{Sample}
	In a search for the most extreme, and thus rare, star-forming galaxies at $z 
> 1$, we exploit the full-sky sub-mm coverage offered by the
the \Planck\ Catalog of Compact Sources (PCCS). The highest frequency observed by \Planck\ (857 GHz / 
350 \micron) contains a dataset of $\sim$ 24,000 point source objects 
\citep{planck13}. From this dataset we limit our searches to point sources at 
Galactic latitude ${\rm |b| > 30\degree}$ to minimize the Galactic source 
contamination. This filtered sample is then cross-correlated with the combined 
catalogs of three \Herschel\ large area surveys: \Herschel\ Multi-tiered 
Extragalactic Survey \citep[HerMES,][]{oliver12}, \Herschel\ Stripe 82 Survey 
\citep[HerS-82,][]{viero14}, and the dedicated \Planck\ follow-up \Herschel\ 
DDT 
``Must-Do" Programme: ``The \Herschel\ and \Planck\ All-Sky Source Snapshot Legacy Survey''. 

%
	
	The details of our selection method can be found in \citet{harrington16} for the \Planck\ - \Herschel\ 
	counterparts with $ S_{\rm 350} \ge 100 
{\rm mJy}$ in our initial follow-up during the Early Science 
Campaign 2 for the \LMT. In brief, we cross-matched \Planck-\Herschel \, counterparts within 150". In total there were 350 \Herschel\ 
counterparts to 56 \Planck\ sources within 150". The higher spatial resolution 
of \Herschel\ allowed us to pinpoint the position of the \Planck\ point sources, enabling follow up studies. 

For 8/11 galaxies observed with the \LMT\
 we detected a single, compact source using the AzTEC 1.1mm camera.  Subsequently 
we detected a strong CO line with the RSR spectrometre.  We make use of the 3 SPIRE bands of \Herschel\ (250, 350, 500 \micron) 
and the additional \LMT\ observations to derive apparent ${\rm \mu L_{IR} > 
10^{13.0-14.5}  L_\odot}$ at ${\rm z_{CO} = 1.33 - 3.26}$ (see Table~\ref{tab:VEG}). The current 
sample in this \GBT\ study consists of observations of only seven of the original eight targets.

\section{\GBT\ Observations}

\begin{table*}
 \caption{Sources and \GBT\ Observations Summary}
\label{tab:GBTrcvr}
 \begin{tabular}{lcccccccc}
  \hline
   Source ID  & RA  & DEC &$\mu L_{IR}^\dagger$ &  \GBT\ RX & Dates & Int. Time (On-Source) 
& ${\rm < T_{sys}} >$ \\
   \hline
   		& J2000 & J2000 &   ($10^{14}  L_\odot)$ & &   2016 & mins & K \\		
  \hline
  PJ142823.9 & 14h28m23.9s  & +35d26m20s & $0.19\pm0.04$&  Q    & 2/12;2/19 &   336 & 100 \\   
  PJ160722.6 &16h07m22.6s  & +73d47m03s &  $0.14\pm0.03$& Q   &2/12;2/19 &   216 & 75 \\  
  \hline
  PJ105353.0  & 10h53m53.0s  & +05d56m21s & $2.9\pm0.4$  &   ${\rm K_{a} }$   &3/30 & 
100 & 40 \\
  PJ112714.5  & 11h27m14.5s  & +42d28m25s & $1.1\pm0.2$  &  ${\rm K_{a} }$  & 3/30 &  
84 & 40 \\
  PJ120207.6  &12h02m07.6s  & +53d34m39s & $1.4\pm0.3$ &  ${\rm K_{a} }$  &3/30 &  80 
& 
40 \\
  PJ132302.9  &  13h23m02.9s  & +55d36m01s &  $1.2\pm0.2$ &  ${\rm K_{a} }$   &3/30 & 
96 & 40 \\   
  PJ160917.8  & 16h09m17.8s  & +60d45m20s & $2.0\pm0.4$  &  ${\rm K_{a} }$   &3/26 &  
92 & 35 \\
  \hline
 \end{tabular}
 
    [1] Q band receiver frequency coverage: $38.2-49.8$GHz. 
    [2] ${\rm K_{a} }$ band receiver frequency coverage: $26.0-40$GHz.
    [3]$^\dagger$  $L_{IR}$ is the far-infrared luminosity integrated between 8-1000 \micron\ .
 
\end{table*}

 	Based on our RSR spectroscopy, two of our 
sources have redshifted CO($1-0$) (i.e. rest-frame ${115.27}$ 
GHz) line emission in the range of the Q band receiver ($38.2-49.8$ GHz) on the 
\GBT. The other five sources fall within the ${\rm K_{a} }$ band receiver ($26.0-40$ 
GHz). We used the low-resolution 1500 MHz bandwidth mode of the backend 
spectrometre, VEGAS. The raw channel resolution corresponds to 1.465 MHz, or 
${\rm \sim 16}$ km s$^{-1}$ in ${\rm K_{a}}$ band, using 1024 channels.
Observations between February 
and March, 2016, took place in typical weather conditions. For both $Q$ and ${\rm K_{a} }$ band observations, we used a
SubBeamNod procedure, nodding the 8m \GBT \, sub-reflector every 
6 seconds between each receiver feed for an integration time of four minutes. 
In 
most cases, this 4 minute procedure was repeated back-to-back for up to an hour 
to achieve the ON source integration times presented in Table ~\ref{tab:GBTrcvr}. 
The atmosphere becomes highly variable at the frequencies within $Q$ and ${\rm K_{a} }$ band, therefore we observed pointing sources roughly 
every hour. The routine pointing and focus procedures allowed us to assess the best azimuth 
and elevation corrections, as well as the best focus values for the peak line 
strength measurements.  

After total-power switching for the standard ${\rm (ON-OFF)/OFF}$ 
GBTIDL 
calculations, we include the observatory's atmospheric model, which tracks 
zenith 
opacity as a function of frequency and time. Each scan is corrected for the 
atmospheric time and frequency variations on the sky, given zenith opacity 
${\rm \tau_{sky} }$ and elevation, EL: ${\rm T_{antenna} = T_{sys} \times 
e^{\frac{\tau_{sky}}{sin(EL)}} \times \frac{(ON-OFF)}{OFF} }$. The elevation 
ranges for Q and ${\rm K_{a}}$ band spanned $33-84 \deg$ and $28-73 \deg$ , 
respectively. The typical ${\rm T_{sys}}$ values ranged from 67-134 K in Q 
band, 
and 30 - 45 K in ${\rm K_{a}}$ band. To convert the measured antenna 
temperature 
in $T_A^{*}$ to flux density we use the calibration factor derived for \GBT: Q 
band scales as ${\rm 1 K  / 1.7}$ Jy, ${\rm K_{a}}$  band as ${\rm 1 K / 1.6 
Jy}$.
	
\begin{figure*}
\begin{center}
\includegraphics[width=15.0cm]{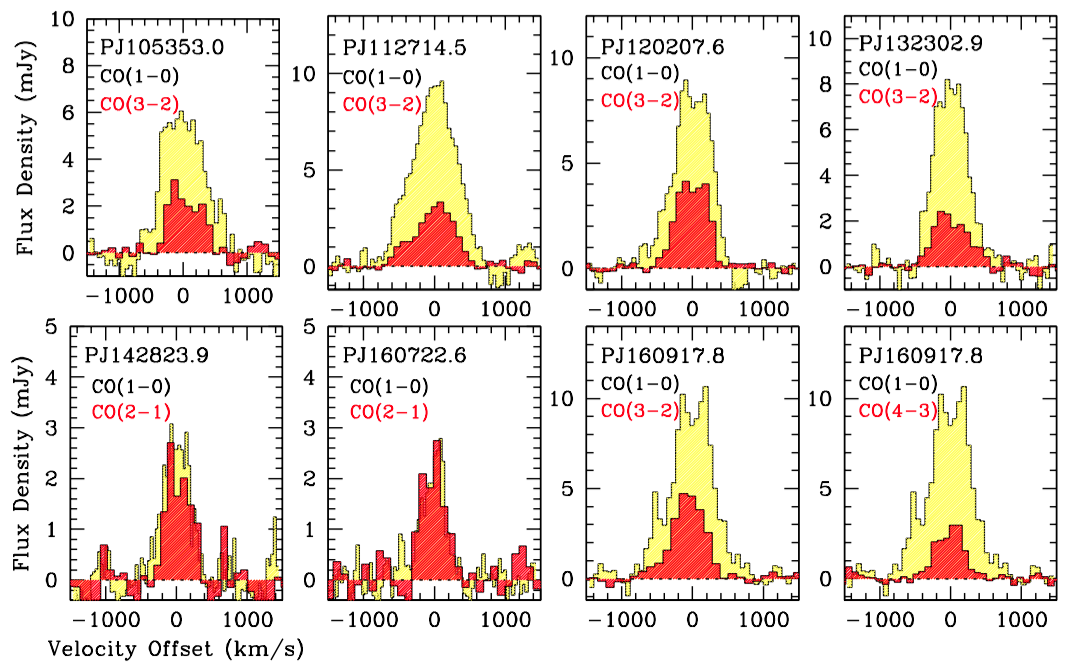}
\end{center}
\vspace*{-5mm}
\caption{The RSR CO spectra (yellow) for all 7 galaxies 
\citep{harrington16} are scaled by ${\rm J_{\rm up}^{2}}$ and overlaid (red) onto the GBT 
CO spectra (yellow) in this study. The comparable line widths and spectral features are coincident. PJ160918 has both its CO($4-3$) and 
($3-2$) lines compared to the ($1-0$) line emission.}
\label{fig:LMTGBT}
\end{figure*}

	 We used a high-pass filter to remove very low-frequency ripples in the 
overall baseline without removing the line emission. The width of the 
high-pass filter was at least twice the expected full-width at half 
maximum (FWHM) of the CO line based on our \LMT\ RSR spectra. We utilised Gaussian smoothing to 
decrease the resolution by a factor of 4, resulting in a 5.86 MHz (${\rm \sim 
50 
\, km s^{-1} }$ for ${\rm K_{a}}$) channel resolution. In this smoothing step, each channel was treated as
independent to avoid correlations in the noise of neighboring channels. As the high-pass filter 
removes the low-frequency ripples, and not the mid-frequency baseline ripples, 
we then fit and remove a baseline (${\rm n_{poly} = 2-3}$) to the emission free 
regions of the spectra.
The resulting spectra can be seen in Fig. \ref{fig:LMTGBT}. We adopt a 30\% total uncertainty given a 
15-20\% flux calibration error, typical 5-10\% pointing/focus drifts and atmospheric losses and a conservative 10-15\% for the 
baseline removal due to the variations across the bandpass at the observed frequencies.

\section{Results: CO ($1-0$) Line Properties}

	We detect 
CO($1-0$) at ${\rm S_{\rm peak}/N_{\rm channel} > 7}$ from each of our seven targets at the expected 
redshifts. We first derive the observed central frequency by fitting a single Gaussian to the CO($1-0$) line emission, confirming the exact
redshifts of these \Planck-\Herschel\ identified galaxies, which had previously been derived using only one CO line from the \LMT (Tab.~\ref{tab:VEG}). 
The spectroscopic 
redshifts span from $1.33 < z < 3.26$. Our new \GBT\ 
measurements further support our previous redshift determinations from the combination of panchromatic photometry (WISE-11 and 22 \micron, 
\Herschel\ SPIRE 250, 350, and 500 \micron, AzTEC 1100 \micron\ and NVSS/FIRST 
radio) and single CO line observations (see Appendix A of \citep{harrington16}).

	We find that the 
CO($1-0$) lines show nearly identical profiles and widths as the 
J$_{\rm up} = 2-4$ CO lines, with FWHM = [375--740 \kms] (see 
Fig.~\ref{fig:LMTGBT}).
It is unlikely that there is a significant amount of gas that excites the CO($1-0$) but 
not, e.g. the CO($3-2$). Therefore, the similar line emission FWHM and line profiles suggests that both transitions are tracing co-spatial volumes. 
	
	We calculated the line luminosity, ${\rm L'_{CO(1-0)}}$, using Eq.~(3) by 
\citet{solomon97},$$L'_{\rm CO} = 3.25\times10^7 S_{\rm CO}\Delta V \nu_{\rm 
obs}^{-2} D_{\rm L}^2(1+z)^{-3}$$ with $S_{\rm CO}\Delta V$ in Jy km s$^{-1}$, 
$\nu_{\rm obs}$ in GHz, and $D_{\rm L}$ in Mpc (see Table~\ref{tab:GBTCO}). 
Since the CO lines are not exactly Gaussian, we also integrated the spectra within 
${\rm \pm 1500}$ \kms to compute $S_{\rm CO}\Delta V$. The corresponding 
Gaussian derived values of $S_{\rm CO}\Delta V$ are the same within 1-sigma. 
Some of the measured apparent $L'_{\rm CO}$ are amongst the brightest, if 
not 
the brightest, for all $z > 1$ QSO, SMG, LBG, including the SPT DSFGs 
\citep{aravena16}, as well as $ z < 1$ ULIRGs, nearby spirals, and low-$z$ 
starburst galaxies \citep{carilli13}. These bright apparent luminosities suggest that our 
galaxies have been strongly magnified. 

\section{Discussion}
\subsection{CO Spectral Line Energy Distributions (SLEDs)}
		
\begin{figure*}
\includegraphics[width=15.0cm]{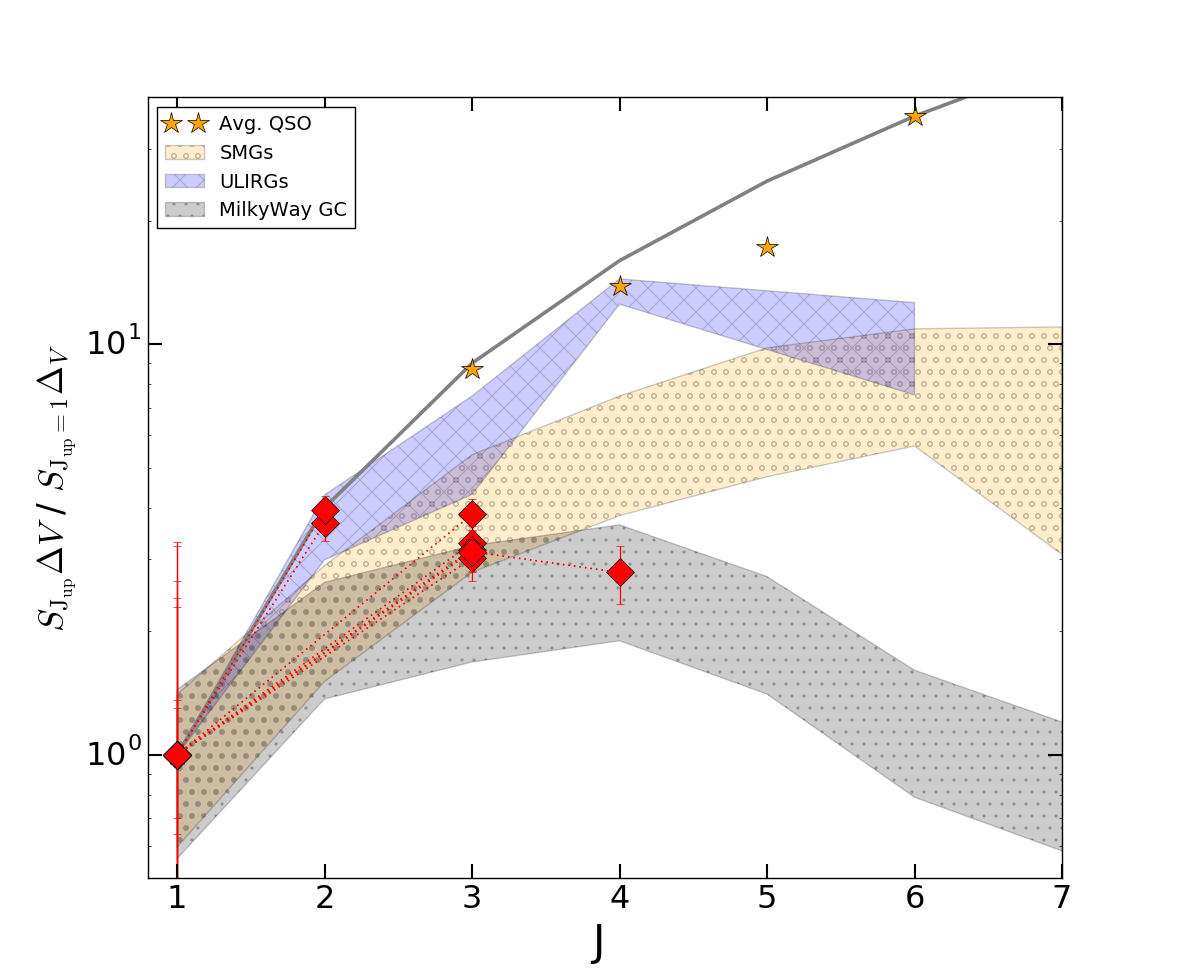}
\caption{Here we plot the velocity-integrated line intensity ratios of J$_{\rm up}/$J$ = 1$,  
normalised 
to the CO($1-0$) derived integrated flux for the current sample. Our seven galaxies (red diamonds) 
are within the spread for average SMGs \citep{bothwell13} (yellow), and can be compared to the low-$z$ (U)LIRG population \citep{pap12} (blue), and the Milky 
Way center \citep{fixsen99} (gray). All regions contain the dispersion between 
the 25th and 75th percentile of the distribution of the CO($1-0$) normalised 
integrated flux. Yellow stars show the average QSO values out to J $= 6$ \citep{carilli13}.}
\label{fig:SratJ}
\end{figure*}
	
	In Fig~\ref{fig:SratJ} we plot the ratio of the line integrated intensity of the higher-J CO $S_{\rm 
CO}\Delta V$ to our CO($1-0$) $S_{\rm CO}\Delta V$. All of our galaxies show sub-thermalised 
excitation conditions. Up to $J \le 3$, we find these values to be consistent 
with both the lower end of the SMG excitation distribution \citep{bothwell13,carilli13} and the upper end of the MW \citep{fixsen99}. The 
uncertainty of the MW measurements and the physical intrinsic SMG dispersion overlap for $J \le 
4$. 
Without higher-J CO lines, where SMGs and the MW differ strongly, it is a 
challenge to disentangle which of these two ISM conditions dominate our galaxies. 
	
We parametrized these CO SLEDs in terms of brightness temperature, or CO line 
luminosity, ratios, $$r_{up,1} = \frac{T_{B} (J_{upper} > 1)}{ T_{B} (1-0)} = 
\frac{L'_{CO} (J_{ upper} > 1)}{L'_{CO} (1 -0)}.$$ For two sources
with only $r_{21}$ we found $< r_{21} > = 0.92$, while the remaining five sources have $< r_{31} > = 0.34$. 
Finally, for the one source with a CO($4-3$) line observation, we found $r_{41} = 0.18$, similar to what has been reported in \citet{hainline06} for an SMG of similar redshift 
(i.e. $z \sim 3.3-3.5$). As noted in
\citep{frayer11}, there is a wide-range in the observed $r_{31} = 0.1 - 1$ in the local starburst 
population. 
The fact that our galaxies fall in the lower end of the SMG excitation distribution, while being some of the most luminous sources
currently known, suggests that they may not be exceptional SMGs, but more strongly magnified sources.
	
	In \citet{harrington16} we showed that all of these sources fall within 
the parameter space for SF powered luminosity (rather than AGN) in a mid-IR to 
far-IR color-color diagnostic plot \citep{kirkpatrick13}. Using their CO SLED we can
further rule out the presence of a powerful QSO in our 
galaxies, as typical QSO host galaxies with powerful AGN activity often exhibit thermalised line ratios out 
to 
${\rm CO(4-3)}$ \citep[e.g][]{riechers06,weiss07}. However, we caution that most QSO hosts with
a good coverage in the CO SLED are strongly lensed objects selected in the
optical/NIR. This may result in a bias towards the excitation conditions within the central region.
\citet{sharon16} show there is a statistical similarity between the $r_{31}$ 
values reported for SMGs and AGN in their sample. However, the line ratios in their sample 
have a global average (AGN and SMG) 3 times higher for $r_{31}$ (in fact close 
to thermalised: $<r_{31}> = 0.9$)
as compared to 
our sources. This suggests that their sample might consist of hybrid SMG/AGN galaxies. 
Our CO SLEDs are currently limited out to J $ = 3$ or $4$, therefore we 
cannot rule out the presence of an AGN. 

\begin{table*}
 \caption{Summary of CO($1-0$) VEGAS Measurements}
 \label{tab:VEG}
 \begin{tabular}{@{}lccccccc}
  \hline
   ID  & $\nu_{obs}$ & $z_{CO(1-0)}$ & $\Delta V$  & $S\Delta V^\dagger$ & $S_{\nu}$ & $\mu L'_{CO}$
\\
    & (GHz) & & (km/s) & (Jy km/s) & Peak (mJy) & ($10^{10}$ K \kms pc$^{2}$) \\
 \hline
\\
  PJ105353.0 & 28.7712  & $3.0053 \pm 0.00016$ & $ 738\pm 38 $ & $4.3\pm1.3$ 
&$6.2\pm1.9$  & $170\pm60$ \\
  PJ112714.5 & 35.6248 & $2.2352\pm0.00006$ & $736\pm20$ & $7.4\pm2.2$ & 
$9.3\pm2.8$ & $178\pm63$ \\
  PJ120207.6 & 33.4970 & $2.442\pm0.00007$ & $602\pm21$ & $5.5\pm1.7$ &  
$8.9\pm2.7$  & $154\pm55$\\
  PJ132302.9 & 33.7350 & $2.4165\pm0.00006$ & $540\pm17$ & $4.7\pm1.4$ &  
$8.4\pm2.5$ &$ 129\pm46$ \\
  PJ142823.9 & 49.5766 & $1.3254\pm0.00005$ & $436 \pm25$ & $1.2\pm0.4$ & 
$2.9\pm0.9$ & $11\pm4$  \\
  PJ160722.6  & 46.4115 & $1.4838\pm0.00006$ & $374 \pm27$ & $1.0\pm0.3$ &  
$2.4\pm 0.7$ & $12\pm4$ \\
  PJ160917.8 & 27.0911 & $3.2567\pm0.00014$ & $705 \pm 31$ & $7.6\pm2.3$ & 
$9.8\pm2.9$  & $343\pm121$\\       
 \hline
 \end{tabular}

[1] $T_A^{*}$ to flux density using the \GBT\ : Q band 
is 1K/1.7 Jy, Ka band is 1K/1.6 Jy.
[2] The reported redshifts correspond to the values obtained after 
velocity offset corrections. The line widths reported indicate the FWHM values 
after correcting for the instrumental resolution. This correction was on 
average 
less than 1\% due to large observed line widths. 
[3] $^\dagger$ The integrated value obtained within the interval of ${\rm 
\pm 1500}$ \kms from the center.

 \end{table*}

\begin{table*}
\caption{Gas Properties}
\label{tab:GBTCO}
\begin{tabular}{@{}lccccccc}
\hline
ID & $L(IR) / L'_{\rm CO(1-0)}$ & $r_{\rm up,1}$  & $\mu M_{\rm ISM}$ & $\mu M_{\rm H2}$ & 
$\tau_{\rm depl}$-CO & $\tau_{\rm depl}$-ISM \\
 & $ L_{\odot} / {\rm (K km s^{-1} pc^{2})}$ &  & ($10^{10} {\rm M_\odot}$) & 
$(\alpha_{\rm CO} 4.3,0.8)$($10^{10}$ ${\rm M_\odot}$)& $ (\alpha_{\rm CO} 4.3,0.8)$ (Myr) & (Myr) \\
\hline
 PJ105353.0 & $ 170\pm65 $ & $0.36\pm0.13 [r_{31}] $ & $624\pm156$ & 
$[732,136]\pm[259, 54]$ & $[239,44]$ &203 \\ 
 PJ112714.5 & $ 62\pm25$ & $0.29\pm0.10 [r_{31}] $ &$160\pm40$ & 
$[764,142]\pm[270, 27]$ & $[636,118]$  & 133 \\ 
 PJ120207.6 & $91\pm38$ & $0.4\pm0.14 [r_{31}] $ & $452\pm113$ & $ 
[663,123]\pm[234, 30]$& $[442,82]$   & 302 \\
 PJ132302.9 & $93\pm36$ & $0.31\pm0.11 [r_{31}] $ &  $215\pm54$ & 
$[557,103]\pm[196, 23]$& $[425,79]$  & 164  \\
 PJ142823.9 & $168\pm69$ & $0.88\pm0.36 [r_{21}] $ & $97\pm24$ & $[49,9]\pm[17, 4]$ & 
$[228,42]$  & 452 \\
 PJ160722.6 & $121\pm50$ & $0.95\pm0.38 [r_{21}] $ & $52\pm13$ & $[50,9]\pm[18, 5]$& 
$[331,62]$  &344\\
 PJ160917.8 & $58\pm24$ & $0.35\pm 0.13 [r_{31}] $ & $465\pm116$ & 
$[1473,274]\pm[521, 85]$& $[694,129]$  & 219  \\       
 PJ160917.8 & $-$ & $0.18\pm0.06 [r_{41}] $  & $-$& $-$ & $-$  & $-$\\       
\hline
\end{tabular}

[1] Unknown lensing amplification $\mu$ is reflected in the derived CO 
luminosity and \H2\ mass. ISM masses were calculated following \citet{scoville16}, scaled to our AzTEC 1.1mm photometry with a fixed 
dust temperature of 25 K.
\end{table*}

\subsection{Ratio of IR Luminosity to CO Line Luminosity}
	The observable ${\rm \mu L_{IR} / \mu L'_{CO(1-0)}}$ ratio 
serves as a proxy for SF efficiency (SFE), and stands independent of the 
unknown 
magnification factor \footnote{We assume, without high angular resolution imaging 
of the two luminosity sources, that the CO($1-0$) emitting region and the 
pervasive dust content are on average co-spatial.}.
The integrated IR emission (8-1000$\micron$) reflects the bulk star-forming activity, while the CO line luminosity indicates the 
amount of gas supplying the ongoing star formation.
		
\begin{figure*}
\includegraphics[width=15.0cm]{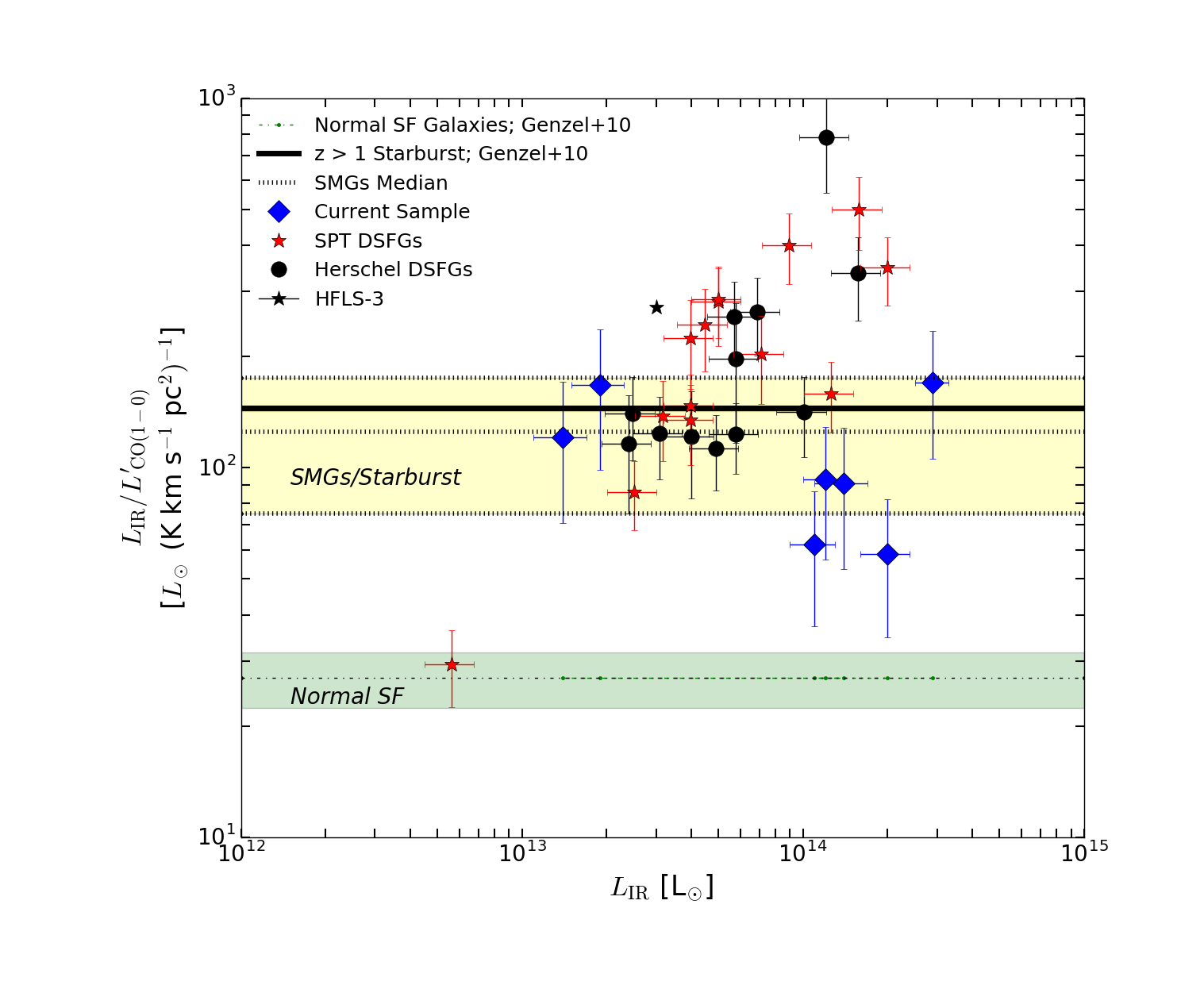}
\caption{Here we present the ${\rm L_{IR} / L'_{CO(1-0)}}$ ratios of our sample 
compared with known, lensed \Herschel\ and SPT DSFGs \citep{harris12, aravena16}, the highly excited HFLS-3 \citep{riechers13}
and the median for all SMGs 
(${\rm 125 \pm 50 L_\odot / K km s^{-1} pc^{-2} }$) compiled in the literature by \citet{frayer11} (shaded 
yellow). We plot 2$\sigma$ boundaries taken from \citet{genzel10} for starburst (140 ${\rm L_{\odot} / K km s^{-1} 
pc^{-2}}$) and typical star-forming galaxies (30 
${\rm L_{\odot} / K \, km s^{-1} pc^{-2}}$). The average for our seven targets 
in this study is ${\rm 110 \pm 22 \, L_{\odot} / (K km s^{-1} pc^{2})}$. }
\label{fig:lirlco}
\end{figure*}

Using the value of this ratio we place 
our 
sample in the context of SB versus typical star-forming galaxies at 
different $z$, IR and CO line luminosity 
\citep[Fig~\ref{fig:lirlco}; e.g.][]{genzel10}. We measure the ${\rm \mu L_{IR} / \mu L'_{CO(1-0)}}$ ratio 
as ${\rm (58-170) \, L_\odot / (K \, km 
s^{-1} 
pc^{2})}$, with ${\rm < 110 \pm 22 > \, {\rm L_\odot} / K}$ \kms pc$^{-2}$ (see 
Fig~\ref{fig:lirlco}). The average value of our galaxies is closer to 140 ${\rm 
L_\odot / (K km s^{-1} pc^{2})}$ observed in SB galaxies, rather than 30 
${\rm L_\odot / (K km s^{-1} pc^{2})}$ observed in typical star-forming galaxies 
\citep{sv05, genzel10, frayer11}. From this we conclude that the $L_{IR} / 
L'_{CO(1-0)}$ values obtained for this subset of \Planck-\Herschel\ sources 
have 
enhanced ratios with respect to typical star-forming galaxies, as expected from their
large apparent $L_{\rm IR}$ \citep{canameras15, harrington16}. 

We note that our sample 
exhibits slightly lower ratios on average compared to both the highly excited system, HFLS-3 \citep[$z \sim 6$][]{riechers13}, as well 
as the lensed SPT DSFGs 
\citep{aravena16} (Fig~\ref{fig:lirlco}). Roughly half of the strongly lensed, dusty \Herschel\ galaxies 
\citep{harris12} are consistent with our sample and lie within the yellow shaded region for SB systems. In contrast, more than half of the
SPT sources have excess $L_{IR}$-to-$L'_{CO(1-0)}$ \citet{aravena16}, although the dispersion is similar for both H-ATLAS and SPT samples.
Our seven \Planck-\Herschel\ sample are unusual as they lack a similar dispersion. This may reflect the ability of the all-sky sub-mm sensitivy and coverage of the \Planck\
survey in detecting the rare galaxies that are, on average, more strongly lensed than the similarly selected H-ATLAS sample.

While the SPT lensed 
galaxies are a similar population at high-$z$ with comparable $L_{IR}$, due to different selection methods 
(350 \micron\ versus mm), the average redshift of their sample is significantly shifted towards a higher value compared to our sample: $< z > = 
3.9$ and $< z > = 
2.3$, respectively. At such a high redshift, $z \sim 4$, a MW type galaxy would be subject to non-negligible 
dust heating due to the CMB \citep{dacunha13}, and may
contribute to the higher $L_{IR}$-to-$L'_{CO(1-0)}$ values observed in the SPT sample. 
At $z \sim 4$, the CMB temperature is also a
sufficient background to radiatively excite the cool reservoirs of CO, particularly the J$= 1 \rightarrow 0$ ground state 
rotational transition, resulting in a dimming of the observed CO line emission. Because (sub)-mm flux measurements are made against the CMB, 
the contrast in the CO ($1-0$) line integrated intensity
via collisional excitation (typically
with H$_{\rm 2}$ molecules), 
compared with the radiatively excitated CO gas from the
CMB background becomes more severe beyond $z = 4$. About 80\% of the 
CO ($1-0$) emission can be recovered against the CMB at $z = 2-3$, but only 50-60\% just beyond $z = 4$ if there would be a
gas kinetic temperature of 40 K \citep{dacunha13}. 

We caution that the effects of the CMB alone cannot account for the differences observed in these luminosity-luminosity ratios, as
the H-ATLAS and SPT sample have a similar spread in their $L_{IR}$-to-$L'_{CO(1-0)}$ values. 
The similar redshift range of the 12 H-ATLAS sample compared to the sample of 7 \Planck-\Herschel\ galaxies in this 
study reveals that the CMB effects cannot explain this offset. The excitation conditions of a multi-phase, multiple gas 
component ISM are also expected to change for each galaxy. One would expect that the density and 
kinetic temperature of the CO ($1-0$) emitting gas (and the gradients across the galaxy) to factor into the 
total attenuation of the CO ($1-0$) line
 emission \citep{tunnard16,tunnard17} and any self-shielding. 
 As the intense star-forming conditions during the redshifts indicated in these three samples
 (SPT, H-ATLAS, \Planck-\Herschel) will give rise to a dynamic set of 
 ISM conditions, these varying gas excitation conditions will
 therefore have non-negligible effects in the observed $L_{IR}$-to-$L'_{CO(1-0)}$ values.
 
\subsection{Total gas mass from $L'_{CO(1-0)}$}
	CO is the second most abundant molecule in the ISM after the highly 
abundant molecular hydrogen, ${\rm H_{2}}$, and the CO($1-0$) line emission is the most direct proxy for 
${\rm H_{2}}$ as it traces even the most diffuse gas. 
Our galaxy integrated CO($1-0$)
line luminosity is converted to a total molecular gas mass assuming an 
$\alpha_{\rm CO}$ conversion factor \citep[see review by][]{bolatto13}. It is 
common 
to use a standard ULIRG conversion, i.e. $\alpha_{\rm CO} = 0.8$, for star 
bursting SMG/DSFGs at high-$z$, although we reference a standard Galactic value in Table~\ref{tab:GBTCO}. 
The similarity of the ${\rm L_{IR} / L'_{CO(1-0)}}$ ratios 
observed in our sample and those of local ULIRGs seems to 
further support the use of a starburst-like $\alpha_{\rm CO}$ conversion factor, even if the centrally compact, concentrated nuclei in 
local ULIRGs may not be representative of the entire ISM environments 
in our galaxies.
We found $\mu M_{\rm H2} 
= (0.9-27.4) \times 10^{11} {\rm M_\odot (\alpha_{\rm CO}/0.8)} $, which are amongst the 
largest apparent gas contents measured at high-$z$, even if a lensed magnification of an order of magnitude
is taken into account \citep[see][]{carilli13}.


We also compare our CO-based gas mass to the ISM gas mass estimates using the empirical calibration from measured
rest-frame dust continuum \citep[e.g.][]{scoville16, scoville17}. 
Using our AzTEC 1.1mm photometry ($\nu_{\rm obs} = 272$ GHz $\rightarrow$ rest-frame 250-470$\micron$), we compute the ISM mass using Eq. 14 of \citet{scoville17}.
The ISM masses we report scatter predictably around the values obtained from a ULIRG or Galactic conversion factor, suggesting that the
value for $\alpha_{\rm CO}$ varies intrinsically from galaxy to galaxy. Later in \S~5.5 we will revist this empirical calibration to compare the CO line 
luminosity to the specific luminosity at rest-frame 850\micron. 

\begin{figure*}
\includegraphics[width=15.0cm]{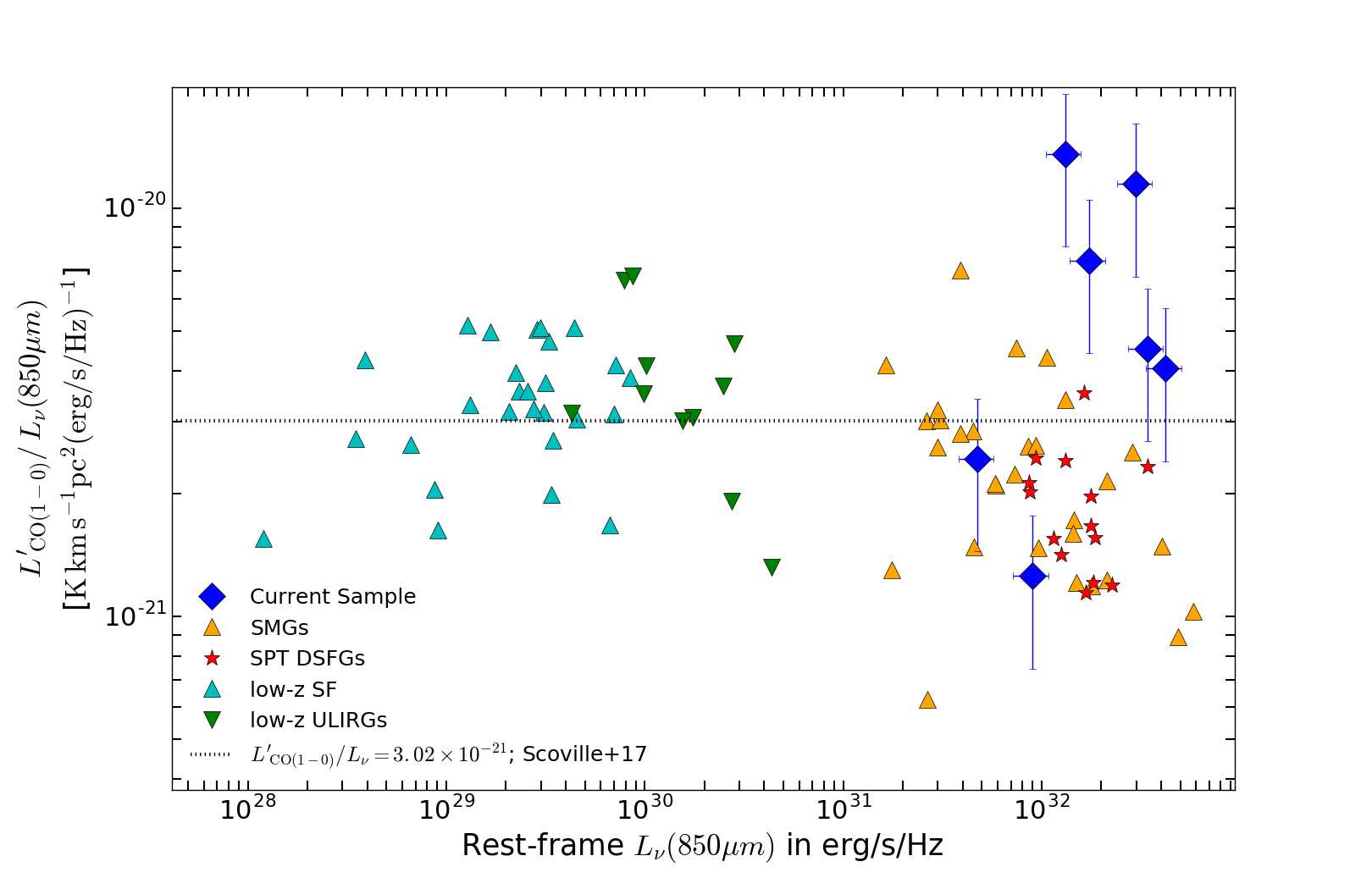}
\caption{We compare our measured $L'_{CO(1-0)}$ to rest-frame $L_{\rm \nu}(353 GHz/850\micron\ )$ in our sample to 
the low-$z$ star-forming galaxies \citep{dale12, young95}, local ULIRGs \citep{mentuch12, chu17, sanders89, sanders91,
solomon97}, $z \sim 2$ SMGs \citep{greve03, harris10, ivison11, carilli11, harris12, riechers11a, lestrade11,thomson12, ivison13, fu13, aravena13,thomson15}, and lensed SPT galaxies \citep{aravena16} with global measurements of CO($1-0$) -- or CO($2-1$) for some SPT sources, where we used $r_{21}= 0.9$ when applicable.
We overplot the best fit linear relation from \citet{scoville17}: $L'_{\rm CO(1-0)} = 3.02\times10^{-21} L_{\rm \nu 850}$. 
}
\label{fig:LCOL850}
\end{figure*}

\subsection{Gas Depletion Time Scales}					
	The amount of time for a galaxy to consume its total molecular gas, 
given its current galaxy integrated star formation rate, is its so-called
depletion time, or gas consumption time scale, $\tau_{\rm depl} = \mu M_{H2} / \mu 
SFR$. This inverse SFE reflects the nature of the SF activity of a galaxy, and is a 
measure that stands independent of the magnification factor in the same way for 
the ${\rm L_{IR}}$-to-CO(${\rm 1-0}$) line luminosity ratios above. 

  To derive our SFR 
estimates we used the integrated 8-1000$\micron$ SED and the empirical calibration \citep{kennicutt98} to convert L$_{\rm IR}$ to SFR --
adopting a Kroupa IMF. The values
we obtain are, uncorrected for magnification amplification, ${\rm \sim 
1500 - 30700 \, M_\odot yr^{-1}}$ \citep{harrington16}.
Combined with the CO-based gas masses reported in \S~5.3, this suggests a depletion time scale of $\tau_{\rm depl} \sim 80 \, {\rm Myr}$
These actively evolving galaxies 
represent a special mode of rapid starburst activity. This is consistent with short gas 
depletion times
observed on the order 
of $\tau_{\rm depl} = 10-100 \, {\rm Myr}$ \citep[e.g.][]{genzel15,bethermin16,aravena16,scoville16}, and also with typical
galaxy-galaxy crossing time ($\sim 100$Myr; \citep{scoville16}).
The rapid $\tau_{\rm depl}$ in these galaxies at 
high-$z$ rival the $\tau_{\rm depl} = 2.2$ Gyr timescales for normal star-forming galaxies at 
$z=0$ 
\citep{leroy13}. 

\subsection{Global Gas to Dust Comparison}
The ratio of measured $L'_{CO(1-0)}$ to rest-frame
specific luminosity at 850\micron\ serves as a foundation for converting the optically thin 
Rayleigh-Jeans dust continuum, observed in the (sub)-mm, into total ISM mass \citep{scoville14, scoville16, scoville17}. 
To infer the rest-frame 850\micron\ of our galaxies, and thus to compare them to the empirical relation,
we use the far-IR SED model fit procedure described by \citet{harrington16}, fitting the \Herschel\ SPIRE 250-500 \micron\ and 
AzTEC 1.1mm photometry with a modified blackbody \citep[Eq. 14][]{yun02} (Fig.~\ref{fig:LCOL850}). 
Several of $z \sim$ 2-3 galaxies lie above the empirical calibration obtained by \citet{scoville17}. In \citet{scoville17} the 
SED analyses was redone using the published sub-mm photometry and CO($1-0$) line emission for the 
30 normal low-$z$ star-forming galaxies \citep{dale12, young95}, 12 low-$z$ ULIRGs \citep{mentuch12, chu17, sanders89, sanders91,
solomon97}, and 30 SMGs 
\citep{greve03, harris10, ivison11, carilli11, harris12, riechers11a, lestrade11,thomson12, ivison13, fu13, aravena13,thomson15} 
at comparable redshifts to our sample. This empirical relation, based primarily on galaxies with solar metallicities, 
was recently validated using $\sim$ 70 main-sequence, low-$z$ star-forming galaxies \citep{hughes17}. 
Without optical or FIR fine-structure emission lines we cannot directly constrain the metallicities
of our sample. However, we do not expect these galaxies to have sub-solar metallicities given their large apparent dust masses ($ \mu M_{d} = [0.1-2] 
\times 10^{10} {\rm M_{\odot} }$) and given the empirical mass-metallicity relationship out to high-$z$
\citep{geach11,saintonge16}.

The SMG/DSFG population predominantly falls below the 1:1 relation, making our small 
sample the first to populate the upper envelope--which corresponds to a higher amount of observed CO gas per unit 850\micron\ dust emission. The highest value of 
$L'_{\rm CO} / L_{\rm 850}$ observed in the SMG population compiled by \citet{scoville17} is the 350\micron\ selected source in \citet{ivison11}. Two of our galaxies are above the 
observed scatter, while three of our galaxies exhibit extreme CO luminosities compared to their rest-frame dust luminosity.
A larger sample is undoubtedly required to further unveil if, as suggested by our sample, there is a larger intrinsic scatter at high-$z$, particularly at  
$\log(L'_{CO(1-0)}) > 10.5 {\rm \, K \, km \,s^{-1} \, pc^{2}}$ and $\log(L_{\rm \nu 850}) > 31.5 \, {\rm erg s^{-1} Hz^{-1}}$.
To compare to the SPT-DSFGs with J$\leq 2$ CO line detections \citep{aravena16}, we take their 18 galaxies with consistent sampling of 
0.25-1.4mm photometry, similar to our 0.25-1.1mm data, and fit their FIR-mm SEDs as described above. 
Those SPT galaxies with only CO($2-1$) were converted to $L'_{\rm CO(1-0)}$ using an $r_{21}= 0.9$.

The relatively high $L'_{\rm CO} / L_{\rm 850}$ ratios observed in our galaxies indicate larger gas-to-dust mass
(GDMR) ratios than observed in previous samples 
(Fig.~\ref{fig:LCOL850}). Converting the AzTEC 1.1mm continuum measurement into dust mass, assuming $T_{\rm d} = 25$K, we found GDMRs in the range [40-200] using the 
CO-based gas mass ($\alpha_{\rm CO} = 0.8$), compared with the average GDMR of $\sim 230$ from the 1.1mm derived ISM mass \citep{scoville17}. This range is both consistent, 
though slightly larger, than observed
in local galaxies with solar metallicities \citep{leroy11a,draine07}. 
Assuming instead a Galactic $\alpha_{\rm CO} = 4.3$, we would infer extremely high GDMRs (up to 1100), only observed in
local, greatly metal-poor dwarf galaxies, e.g., the blue compact dwarf, I Zwicky 18, with 
$1/50$ solar metallacity \citep{annibali15}. The assumption of $\alpha_{\rm CO}$, as well as the choice of dust temperature in the ISM mass calculations
ultimately determines the derived GDMRs. 


		
\section{Conclusions}
Using VEGAS on the \GBT, we have successfully measured the CO($1-0$) line emission for 
seven of the most gas-rich SMGs/DSFGs studied to date. The 
key results of this study can be summarised as: 
\begin{itemize}
 \item We have confirmed the previously determined spectroscopic
redshifts reported by \citet{harrington16}

\item The linewidths/profiles for the low-J RSR and CO($1-0$) VEGAS measurements are nearly 
identical; therefore the emitting regions are likely co-spatial, with ${\rm < FWHM > = 590}$ \kms,

\item  The CO SLEDs of the galaxies in our sample are
indicative of a gas component with sub-thermal excitation conditions: CO line luminosity ratios of $< r_{21} > = 0.92$ (2 sources), $< r_{31} 
> =0.34$ (5 sources), and $r_{41}  = 0.18$ (1 source)

\item  We find enhanced $ L_{\rm IR} / L'_{\rm CO(1-0)}$ ratios with 
respect to normal star-forming systems, as we report an average value of ${\rm 110 
\pm 
22 \, L_\odot / (K km s^{-1} pc^{2})}$, comparable to the median of other well 
studied SMGs.
\end{itemize}
With the CO($1-0$) line emission we place tight constraints on 
the 
total molecular gas mass, and allow future CO SLED analyses to benefit from having the fundamental rotational transition observed.
The large gas masses obtained are $ \mu M_{H2} = 
(0.9 - 27) \times 10^{11} (\alpha_{\rm CO}/0.8) {\rm M_\odot }$. The average gas depletion time we find is $\tau_{\rm depl} \sim 80 {\rm Myr}$. These extremely luminous IR galaxies 
(with $L_{IR} \ge 10^{13-14} L\odot$) exhibit rapid depletion timescales, and 
we 
are likely capturing this light from a relatively short-lived starburst episode.

\section{Acknowledgements}
The quality of this manuscript, and the scientific 
considerations, have been improved significantly after discussions
with N. Scoville, L. Liang and M. Sargent. All authors would like to thank the referee
for their kind considerations of how this paper can be developed so as to make its most useful contribution to the field. 
KCH and DB would like to also thank Karen O'Neil and Toney Minter for 
organizing, and supporting our travel to, the first \GBT\ remote observing school at the Green Bank Telescope.
KCH would
like to express great thanks to D. Frayer and all of the control room operators (specifically
Amber and Donna) for making the observations run as smooth as possible. 
The Green Bank Observatory is a facility of the National Science Foundation operated 
under cooperative agreement with Associated Universities, Inc. KCH and DB heartily acknowledge our
financial support granted to us by UMass Amherst's Astronomy Department, while KCH much appreciated the ongoing support from the 
UMass Amherst Commonwealth Honors College Research Grant which led to the travel to the \GBT\ remote observing school in January, 2016.
KCH, AK and EFJA acknowledge the support by the Collaborative Research Center (CRC) 956, subproject A1, funded by the 
Deutsche Forschungsgemeinschaft (DFG). KCH acknowledges the financial support from the CRC-956 student exchange program. This 
opportunity has enabled KCH to engage in useful discussions and supervision with D. Riechers, who provided strong guidance in the final stages 
of this manuscript. Support for BM was provided by the DFG priority program 1573 ‘The physics of the interstellar medium'. 
 KCH is a member of, and receives financial support for his research by,
the International Max-Planck research school for Astronomy and Astrophysics
at the Universities of Bonn and Cologne and through the Max-Planck-Institut f\"{u}r Radioastronomie. 
\bibliography{allrefs_2}

\end{document}